%% file: TopUpsi.tex
\documentclass[preprint,10pt]{elsarticle}

\usepackage{amssymb}
\usepackage{amsmath}
\usepackage{subcaption}
\RequirePackage{amssymb,amsfonts,amsthm}
\RequirePackage{mathrsfs}
\usepackage[left=30mm, right=45mm]{geometry}

\usepackage{hyperref}
\hypersetup{
colorlinks=true, 
linktoc=all,     
linkcolor=blue,  
}

\newcommand{\Br}{\mathsf{Br}}
\newcommand \mupsi {m_{\Upsilon}}
\newcommand {\calO}{\mathcal {O}}
\newcommand  \mfrac[2]   {\displaystyle \frac{ #1}{#2} }

\journal{Nuclear Physics B}

\begin{document}

\begin{frontmatter}




\title{Top-quark decay into $\Upsilon$-meson }


\author[label1]{S.~Slabospitskii}

\address[label1]{NRC ``Kurchatov Institute'' - IHEP, Protvino, Moscow Region, Russia}
 
\ead{Sergei.Slabospitskii@ihep.ru}

\begin{abstract}
  The calculation of the partial width of the rare $t$-quark decay
  into $\Upsilon$-meson, $W$-boson and $b$-quark ($t \to \Upsilon W b$)
  is presented.  The  branching ratio equals
  $\Br(t \to \Upsilon W b) = 1.3 \times 10^{-5}$
  that make possible searches for
  this rare $t$-quark decay at LHC.
  \\ [5mm]
\end{abstract}


\end{frontmatter}


\include{article_topupsi}

\bibliographystyle{elsarticle-num} 
\biboptions{numbers,sort&compress}
\bibliography{TopUpsi.bib}

\end{document}

%% file: article_topupsi.tex
\section{Introduction}
\label{sec:intro}

In the SM the decay $t \to b W$ is by far the dominant one.
The rates for other decay channels are predicted to be smaller by
several orders of magnitude in the SM~\cite{Beneke:2000hk}.

For example, for  semi-exclusive
$t$-quark decays the interaction of quarks among the $t$-decay
products may lead to final states with one hadron (meson)
recoiling against a jet.
The  decays of the top through an off-shell $W$ with virtual mass
$M_{W^*}$ 
near to some resonance $h$, like $\pi^+, \rho^+$, $K^+$, $D_s^+$,
leads to the estimate as follows~\cite{Beneke:2000hk}:
\begin{eqnarray}
  \Gamma(t \to b h) \approx \mfrac{G_F^2 m_t^3}{144 \pi}  f_M^2 |V_{q q'}|^2
  \label{eq-1}
\end{eqnarray}    
where the parameter $f_M$ is same as a well-known coupling
$f_{\pi}$. 
The typical values of the corresponding branching ratios are too small to
be measured~\cite{Beneke:2000hk}:
\begin{eqnarray}
  \Br(t \to b \pi) \sim 4  \times 10^{-8},
  \quad \Br(t \to b D_s) \sim 2 \times 10^{-7}
  \label{eq-2}
\end{eqnarray}
\noindent 
There are several two-body $t$-quark decay through flavour changing neutral
currents: 
\begin{eqnarray*}
  t \to \gamma \, q, \quad t \to Z \, q,
  \quad t \to g \, q; \qquad q = u, c
\end{eqnarray*}
These processes in the SM can occur due to loop contribution only
and are highly suppressed due
to GIM mechanism. The estimated branching ratios are as
follows~\cite{Beneke:2000hk}:
\begin{eqnarray*}
  \Br(t \to V \, q) \sim \calO(10^{-11} \div 10^{-13}),
  \quad V = \gamma, Z, g, \;\; q= u, c
 \end{eqnarray*} 
In addition, it worth noting that almost all ``interesting''
$t$-quark rare decays have very small branching ratios and
almost impossible to measured in experiment.

Among rare top-quark decays one can single out the processes
with the production of heavy quark $Q \bar{Q}'$-pair (for example,
$b \bar{b} $) followed by the formation of a heavy $M(Q \bar{Q}')$ - meson.
In this case, the description of such mesons production 
allows the use of the
NRQCD-model~\cite{Kartvelishvili:1988pu,Bodwin:1994jh}.

Note, that the  top quarks production processes with  subsequent  
$t$-quark decays  into  heavy quark $Q \bar{Q}'$-pair
is described within SM   with high accuracy. 
Therefore, the search and study of such $t$-quark rare decays 
can allow, in particular, to find out in more detail which models
formation of quarkonium
(Color-Evaporation Model, the Color-Singlet Model
or the Color-Octet Mechanism, see~\cite{Lansberg:2019adr}
for detailed discussion of various mechanisms) describe more correctly
such processes. 

In this article we calculate the $t$-quark decay widths
into $\Upsilon$-meson 
within NRQCD model~\cite{Kartvelishvili:1988pu,Bodwin:1994jh}.
As will be seen below, at least one decay channel has
a ``relatively'' large branching fraction, providing an opportunity
for experimental searches.

\section {The effective $b \, \bar{b} \, \Upsilon$-vertex}
\begin{figure}[h!]
  \begin{center}
\includegraphics[width=0.30\textwidth,clip]{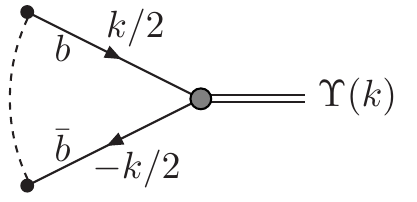} 
 \end{center}
\vspace{-5mm}
  \caption{$b \, \bar{b} \,  \Upsilon$ vertex within NRQCD approach}
     \label{fig:tupsi-1}
\end{figure}

Within NRQCD approach the integration on virtual momentum
 in the loop with two heavy quarks that entered into
heavy $M(Q \bar{Q})$ meson
 (see fig.~\ref{fig:tupsi-1})

\noindent
effectively produces the following expression
(see~\cite{Kartvelishvili:1988pu,Bodwin:1994jh}) for details:
\begin{eqnarray}
  \left.
  \begin{array}{ll}
  &
  \int \mfrac{d^4 p}{\imath (2 \pi)^4} G\left(- \mfrac{k}{2} + p \right)
  \hat{\varepsilon} \,
  G\left(\mfrac{k}{2} +p\right) \{\cdots\} \Psi_{\Upsilon}
\\
 \Rightarrow & 
  \mfrac{R_s(0)}{\sqrt{4 \pi M^3}}
  \left(m_b - \mfrac{\hat{k}}{2}\right) \hat{\varepsilon}
  \left(m_b + \mfrac{\hat{k}}{2}\right)
  \{\cdots\}
  \end{array}
  \right.
    \label{eq-3}
 \end{eqnarray}   
where $G = (m + \hat{p}) / (m^2 - p^2)$ is fermion propagator,
$M = \mupsi$ stands for $\Upsilon$-meson mass,
$\Psi_{\Upsilon}$ is the $\Upsilon$-meson wave function, $\{\cdots\}$
 is other terms in the loop;  
 $\varepsilon^{\mu}$ is the polarization vector of the 
$\Upsilon$-meson.
The  $\Upsilon$-meson wave function at the origin of the $R_S(0)$
is related to the lepton decay
width~\cite{Kartvelishvili:1988pu,Bodwin:1994jh} as follows:
\begin{eqnarray}
  && \Gamma(\Upsilon  \to \ell^+ \ell^-) = \frac{4 e_b^2 \alpha^2}{M^2}
  |R_s(0)|^2
  \label{eq-4}
 \end{eqnarray}
here $e_b$ is $b$-quark charge, $\alpha = e^2 /(4\pi)$.

Note, that in the final expression~(\ref{eq-3})
the  heavy $b$-quarks (entered in the
heavy $\Upsilon$-meson vertex) are considered to be on-shell
with mass equals:
\begin{eqnarray*}
  m_b = \mfrac{M}{2}
  \end{eqnarray*} 
Taking into account that $(k \varepsilon) = 0$ we get the final
expression for $b \bar{b} \Upsilon$ vertex:
\begin{eqnarray}
  &&
  V(b \bar{b}  \Upsilon) = g_{\Upsilon}
   \hat{\varepsilon} (M + \hat{k}),
  \quad  g_{\Upsilon} = 
  \frac{M}{2} \cdot  
  \frac{R_s(0)}{\sqrt{4 \pi M^3}}
  =  \frac{R_s(0)}{4 \sqrt{ \pi M}}
    \label{eq-5}
 \end{eqnarray}


\section {$t \to \Upsilon \, c$ decay}
\begin{figure}[!h]
  \begin{center}
\includegraphics[width=0.40\textwidth,clip]{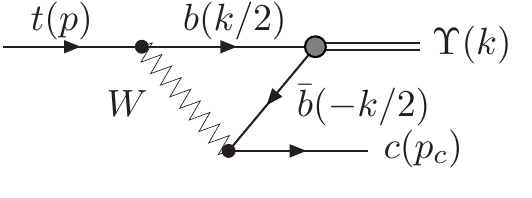} 
 \end{center}
\vspace{-5mm}
  \caption{$t$-quark decay $t \to \Upsilon c$}
     \label{fig:tupsi-2}
\end{figure}

In this section we present the evaluation of the
two-body $t$-quark decay width
\begin{eqnarray}
  && t \to \Upsilon \, c  
   \label{eq-6}
 \end{eqnarray}   
within NRQCD Color Singlet model approach.
This width  was calculated previously
in~\cite{Handoko:1999iu, dEnterria:2020ygk}.
For the sake of completeness
we repeat the evaluation of this quantity.  
The  diagram describing this decay is shown in fig.~\ref{fig:tupsi-2}.
We set the mass of the light $c$-quark equals zero;
 $m_t$ is the mass of $t$-quark, $M = m_{\Upsilon}$ is the
$\Upsilon$ mass{\footnote{
    Throughout of this  article we follow~\cite{Borodulin:2017pwh}
for the notations, the  SM vertices and SM parameters.}}.

\noindent
\noindent 
The amplitude has the following form (see~\cite{Kartvelishvili:1988pu}
for details):
\begin{eqnarray}
  A & =& \left( \frac{g_{\Upsilon} g^2  V_{bc}}{2} \right)
  \, D^{\alpha \beta}_W \;\; \bar{u}(q) \gamma^{\alpha} P_L
   \hat{\varepsilon}
   \left(M  + \hat{k} \right)
   \gamma^{\beta} P_L \, u(p),
   \qquad  P_L = (1 - \gamma^5)/2
   \label{eq-7}
\end{eqnarray}
where $ g = 2  M_W \sqrt{\sqrt{2} G_F}$ ($G_F$ is the Fermi coupling
constant);  $D^{\alpha \beta}_W$ is
the $W$-boson propagator, $\varepsilon$ is the $\Upsilon$-meson
polarization vector:
\begin{eqnarray*} 
  && D^{\alpha \beta}_W = \mfrac{g^{\alpha \beta} -p_W^{\alpha} p_W^{\beta} /M_W^2}
  {p_W^2 - M_W^2},
  \quad    \sum_{pol \Upsilon} \varepsilon^{\mu} \varepsilon^{\nu}
   = g^{\mu \nu} - \frac{k^{\mu} k^{\nu}}{M^2},
   \quad
   (\varepsilon \, k) = 0
   \nonumber
\end{eqnarray*}
\noindent 
 Then the decay width equals
\begin{eqnarray}
   \Gamma(t \to c \, \Upsilon ) &=&
   \frac{\Big(g_{\Upsilon} \, g^2 \, V_{bc}\Big)^2 m_t }
        {768 \, \pi \, \tilde{Z}^2}\left(1 -  \frac{M^2}{m_t^2}\right)^2 \,
\times  U 
  \label{eq-8}
  \\
  U  &=& 6 \mupsi^2 F^2 + (m_t^2 - M^2)
  \left[ 8 \Big(1 + \mfrac{M^2}{8M_W^2}\Big)^2
    -  \mfrac{m_t^2\mupsi^2}{2 M_W^4}
    \right],
  \quad F = \mfrac{p_W^2}{M_W^2} - 2
      \nonumber
 \\
  p_W^2 & = & \mfrac{m_t^2}{2} - \mfrac{M^2}{4},
 \quad
 \tilde{Z}^2 = \left( M_W^2 - p_W^2\right)^2 + \left(\Gamma_W M_W\right)^2
\nonumber      
\end{eqnarray} 
here $\Gamma_W$ is the total decay width of the $W$-boson.
\noindent 
The resulted width (with $m_t = 172.5$~GeV,
$|V_{bc}| = 0.04$~\cite{Zyla:2020zbs}) equals
\begin{eqnarray}
  \Gamma(t \to \Upsilon \, c) &=& 6.35 \times 10^{-10} \;\;
  \hbox{GeV}
   \label{eq-9}
 \end{eqnarray}
and is very similar to previous result~\cite{Handoko:1999iu}.
At the same time this quantity is 2.5 smaller then result
from~\cite{dEnterria:2020ygk}. This difference can be explained by the fact
that authors used contributions of both color singlet and color octet
to this decay channel (see~\cite{dEnterria:2020ygk} for details).

For calculation of the branching ratio we use LO $t$-quark decay width
value of
 \begin{eqnarray}
  \Gamma( t \to b \, W^+) &=& 1.47  \;\;   \hbox{GeV}
   \label{eq-10}
 \end{eqnarray}
 and get the corresponding branching ratio for this decay channel
 \begin{eqnarray}
   \Br(t \to \Upsilon \, c) = 
   \mfrac{\Gamma(t \to \Upsilon \, c)}{\Gamma(t\to b W+)_{LO}}
   &=& 4.32 \times 10^{-10} 
   \label{eq-11}
 \end{eqnarray}

\section{Top-quark decay $t \to \Upsilon \, W \, b$}
It follows from previous section that  two-body $t$-quark decay
$t \to \Upsilon c$  is very small (see~(\ref{eq-11})).
 It is explained by very small value of $|V_{bc}| \approx 0.04$
 and high virtuality of the $W$-boson (
$p_W^2 = m_t^2 / 2 - M^2 /4 \gg M_W^2$, see~(\ref{eq-8})).

To avoid such suppression factors we consider $t$-quark decay
width additional $b \bar{b}$-pair production in the final state:
\begin{eqnarray}
t \to b W^+ \; b \bar{b}
 \label{eq-12}
\end{eqnarray}
This decay process is described by 28 Feynman diagrams.
We use the C++ version of the TopReX package~\cite{Slabospitsky:2002ag}
for calculation the  decay width into this channel. The results  equal
\begin{eqnarray}
  \left.
  \begin{array}{lcl}
  \Gamma(t \to b W^+ b \bar{b}) &=& 8.37 \times 10^{-4}
  \;\;\; \hbox{GeV}
  \\
  \Br(t \to b W^+ b \bar{b}) & =& 5.7 \times 10^{-4}
  \end{array}
  \right.
  \label{eq-13}
   \end{eqnarray}

\begin{figure}[h!]
  \begin{center}
\includegraphics[width=0.80\textwidth,clip]{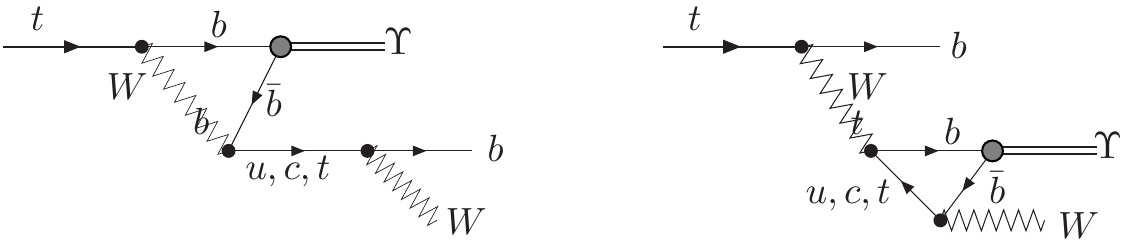} 
 \end{center}
\vspace{-5mm}
\caption{The diagrams describing $t \to \Upsilon W b$ decay
  through  $W$-boson exchange. }
     \label{fig:tupsi-3}
\end{figure}

However, the diagrams with $b \bar{b}$ pair production due to Higgs,
$Z$-boson or $\gamma$ exchange
are highly suppressed (due to small couplings).
As a result we have 4 diagrams, describing $t \to \Upsilon W b$ decay
channel. 
The diagrams with $W$-boson exchange (see
fig.~\ref{fig:tupsi-3})  are also highly suppressed
(due to small couplings and high virtuality of intermediate
$u, c, t$-quarks and $W$).

\begin{figure}[h!]
  \begin{center}
\includegraphics[width=0.95\textwidth,clip]{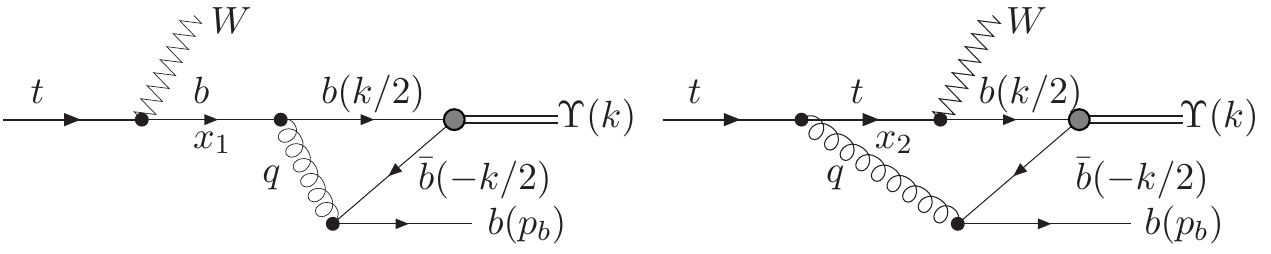} 
 \end{center}
\vspace{-5mm}
  \caption{The diagrams describing $t \to \Upsilon W^+ b$ decay
  through  gluon exchange. }
     \label{fig:tupsi-4}
\end{figure}

\noindent 
Therefore,  the dominant contribution to the amplitude of
$t \to \Upsilon W b$ decay comes from two diagrams
with gluon exchange
(see fig.~\ref{fig:tupsi-4}).

The amplitude $A$ has the form (the particle's momenta notations are
shown in the fig.~\ref{fig:tupsi-4}): 
\begin{eqnarray}
  \left.
  \begin{array}{lcl}
     A &=& A_1 +A_2,
\qquad A_1 = \left( \mfrac{g_{\Upsilon} g g_s^2 }{\sqrt{2} z_1 q^2} \right)
  \, \rho^{\alpha \beta}_{g} \; \times T_1,
  \quad 
  A_2  =  \left( \mfrac{g_{\Upsilon} g g_s^2 }{\sqrt{2} z_2 q^2} \right)
 \rho^{\alpha \beta}_{g} \times T_2
   \\
 T_1 &=& \rho^{\alpha \beta}_{g} \;
\hat{\varepsilon}_W^{\lambda} \; \bar{u}(p_b) \gamma^{\alpha}
     \hat{\varepsilon}_{\Upsilon}
   \left(M + \hat{k}\right)
   \gamma^{\beta} \left(\mfrac{M}{2} + \hat{x}_1 \right)
   \gamma^{\lambda} P_L \,    u(p)
   \\
  T_2 &=& \rho^{\alpha \beta}_{g} \; \hat{\varepsilon}_W^{\lambda}
    \; \bar{u}(p_b) \gamma^{\alpha}
    \hat{\varepsilon}
  \left(M  + \hat{k} \right)
   \gamma^{\lambda} P_L 
    \left(m + \hat{x}_2 \right) \gamma^{\beta}
    \,    u(p)
    \\
    &&
    x_1 = q + k/2, \quad x_2 = p - q, \quad q = p_b + k/2,
   \quad 
   z_1 = \mfrac{M^2}{4} - x_1^2,
  \quad
  z_2 = m_t^2 - x_2^2
  \end{array}
  \right.
      \label{eq-14}
   \end{eqnarray}
where
\begin{eqnarray*}
  \sum_{pol W} \varepsilon_W^{\lambda} \varepsilon_W^{\lambda'} 
   = g^{\lambda  \lambda'} - \frac{p_W^{\lambda} p_W^{\lambda' }}{M_W^2},
   \;\; 
   \sum_{pol \Upsilon} \varepsilon^{\mu} \varepsilon^{\nu}
   = g^{\mu \nu} - \frac{k^{\mu} k^{\nu}}{M^2}
\end{eqnarray*}
The square of the full amplitude is rather cumbersome
and we present it in the Appendix. 

As before we use the the C++ version of the TopReX
package~\cite{Slabospitsky:2002ag} for calculation of 
the  decay width. 
In the  table~\ref{tab:1} we present the results for three $\Upsilon$-meson
states ($\Upsilon(1S), \Upsilon(2S)$ and
$\Upsilon(3S)$).

\begin{table}[h!] 
  \caption{
    The partial top-quark decay widths ($\Gamma$) and
    branching ratios ($\Br$) 
    into three $\Upsilon$-meson states. The widths are in GeV. }
  \label{tab:1}
\begin{center}
 \renewcommand{\arraystretch}{1.2}
 \begin{tabular}{l||c|c||c|c}
    $\Upsilon$  &
 $\Gamma(t \to \Upsilon W^+ b)$ & $\Br(t \to \Upsilon W^+ b)$ 
    & $\Gamma(t \to \Upsilon c)$ & $\Br(t \to \Upsilon c)$   
 \\ \hline
 $\Upsilon(1S)$  & $1.95 \times 10^{-5}$ & $1.33 \times 10^{-5}$
 &  $6.4 \times 10^{-10}$ &  $4.35 \times 10^{-10}$
 \\ \hline 
 $\Upsilon(2S)$  & $0.83 \times 10^{-5}$  & $0.56 \times 10^{-5}$
 & $3.1 \times 10^{-10}$  & $2.11 \times 10^{-10}$ 
 \\ \hline 
 $\Upsilon(3S)$  & $0.58 \times 10^{-5}$  & $0.33 \times 10^{-5}$
 & $2.3 \times 10^{-10}$  & $1.56 \times 10^{-10}$ 
  \end{tabular}
\end{center}
\end{table}
\renewcommand{\arraystretch}{1.}

As mentioned above  two-body
 $t$-quark decays $t \to \Upsilon(nS) c$ (two right columns)
have very small branching ratios for experimental study.
On the other hand the decay channel 
$t \to \Upsilon \, W^+ b$ looks much more promising for
experimental searches.

\begin{figure}[h!]
  \begin{center}
\includegraphics[width=0.75\textwidth,clip]{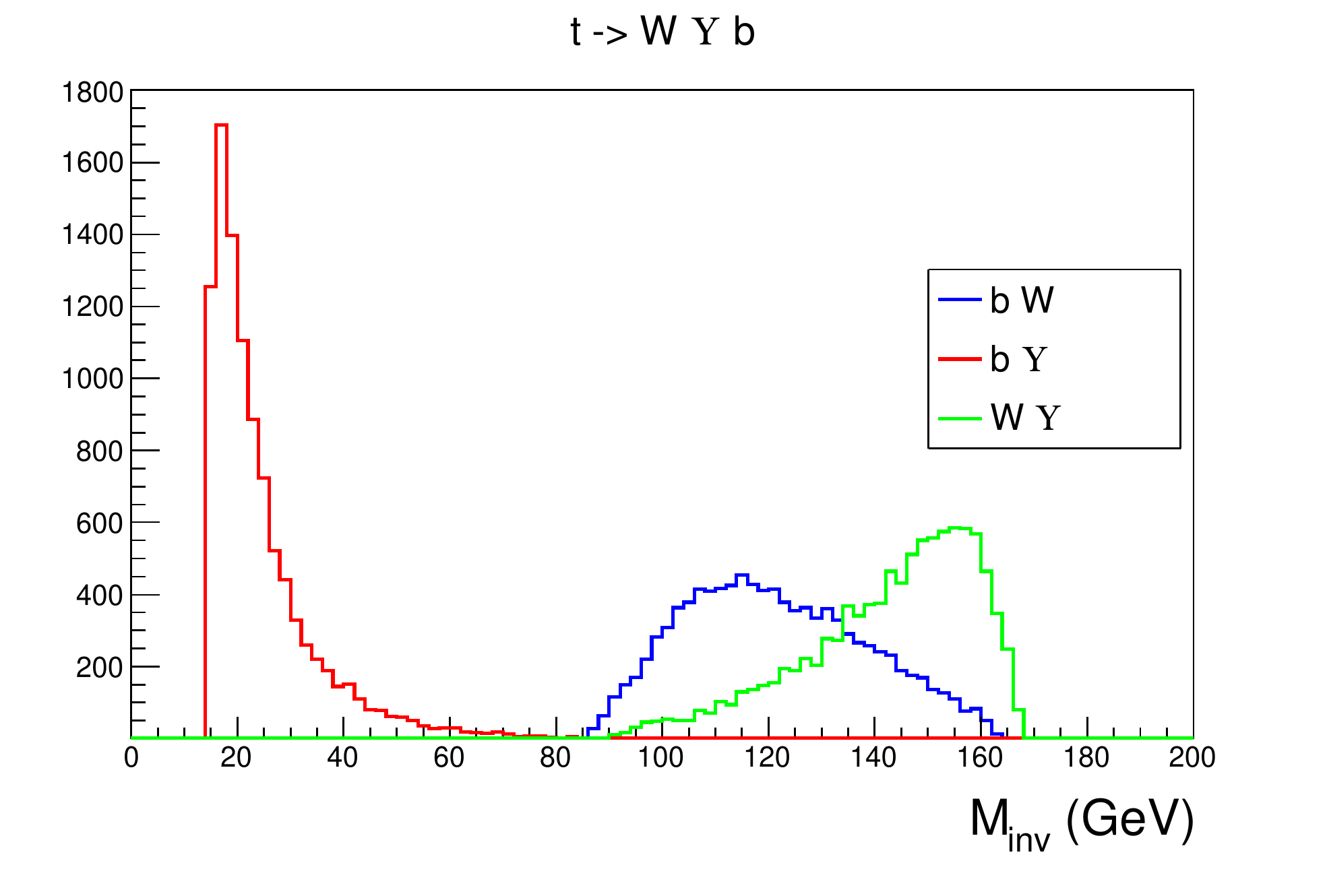} 
 \end{center}
\vspace{-5mm}
\caption{$d \Gamma / d M_{inv}$ distributions.
  The left (red), central (blue),  and right (green)
  curves correspond to $M_{inv} = M(b \, \Upsilon)$,
  $M(b \, W)$ and $M(W \, \Upsilon)$ invariant masses, respectively.
  \label{fig:tupsi-5}
  }
\end{figure}

In the fig.~\ref{fig:tupsi-5} we present the distributions
on invariant masses of the final state particles:
$M(b, \, W)$, 
$M(b, \, \Upsilon)$,  and 
$M(W, \, \Upsilon)$. 

 As it seen the final $W$ and $b$-quark
 are rather well separated, while $b$ and $\Upsilon$ pair
 (the left curve, fig.~\ref{fig:tupsi-5})
has an invariant mass very close to $m_b$ + $\mupsi$. Therefore, one
may expect that the $\Upsilon$-meson will produce
dominantly inside final $b$-jet.


Now we present  very rough estimates of the
expected number of events for this rare $t$-quark decay channel
for LHC Run-2 option. We consider the process of $t \bar{t}$-pair
production with subsequent  $t$-quark (or $\bar{t}$-quark)
decay int three $\Upsilon(nS)$ states $t \to \Upsilon(nS) W b$,
$n = 1, 2 3$. 
The total $t \bar{t}$ cross section, extrapolated to the
full phase space, is~\cite{Grancagnolo:2019uvn}: 
\begin{eqnarray}
  \sigma_{t \bar{t}} = 803 \pm 2 (\hbox{stat}) \pm 25
  (\hbox{syst}) \pm 20(\hbox{lumi}) \;\;\; \hbox{pb}
    \label{eq-15}
\end{eqnarray}
Then, for estimation the expected number of events
we use the following options:
\\
-- the LHC Run-2 integrated luminosity equals $L_{tot} = 100$~fb${}^{-1}$,
\\
-- $W^+ W^-$ decay into lepton and quark pairs
$W^+ W^- \to e(\mu) \, \nu \;\; q \bar{q'}$, \\
-- all three $\Upsilon$ states decay into charged leptons
$\Upsilon(nS) \to ee$ or $\mu \mu$.

As a result, at LHC Run-2 the expected  number
of events for $t \bar{t}$-pair
production with subsequent $t \to \Upsilon \, W^+ b$ decay
are as follows:
\begin{eqnarray}
  && pp \to t \bar{t}, \quad t \to \Upsilon_{1S + 2S +3S} W b,
  \quad \Upsilon \to \ell^+ \ell^-
  \nonumber
  \\
  &&
  \left.
  \begin{array}{lcl}
  N(\Upsilon_{1S+2S+3S}(\to \ell \ell) \,  b \bar{b} \,  W^+ W^- ) &=& 230
  \\
  N(\Upsilon_{1S+2S+3S}(\to \ell \ell) \,  b \bar{b} \, \ell \nu q \bar{q}')
  &=& 80
  \end{array}
  \right.
    \label{eq-16}
\end{eqnarray}

The total number of events $N = 80$ is not very large.
However, this number looks more or less  suitable  for the
experimental study.


\section*{Conclusion}
In this paper  the
 calculation of the partial width
  for rare $t$-quark decay into $\Upsilon$-meson ($t \to \Upsilon W b$)
  is presented.
  The decay width was evaluated within NRQCD-model.
  The calculated branching ratio equals
  $\Br(t \to \Upsilon(1S) W b) = 1.3 \times 10^{-5}$
  that make possible searches for
  this rare $t$-quark decay at LHC.

\vspace{10mm}
\section*{Acknowledgments}
    In conclusion the author is  grateful to V.F.~Obraztsov 
    for multiple and fruitful discussions.

\vspace{2mm}

\appendix
\section*{  Appendix }
Here we present the amplitude square $|T|^2$ from~(\ref{eq-14}). 
The parameters  $m_t, M$, and $M_W$ stand for $t$-quark, $\Upsilon$
and $W$-boson masses, respectively. 
$(a.b)$ is the scalar product of two 4-momenta. 
\begin{eqnarray*}
 |T|^2 & =&  |(T_1 + T_2)|^2  =
  8 \, \left(\chi_1 + \chi_2 + \chi_{int} \right)
\\ 
  \chi_1  & = & 
4(p.p_b)(p_b.k) M^2 + 4(p.p_b)(p_b.k)^2
  - 2(p.p_b) M^4  + 4(p.k)(p_b.k)M^2 - (p.k) M^4
   \\
 &  + &\mfrac{2(p.p_W)}{M_W^2} \Big [ 4(p_b.k)(p_b.p_W)M^2
     + 4(p_b.k)(k.p_W)M^2
     + 4(p_b.k)^2 (p_b.p_W) - 2(p_b.p_W)M^4
     \\
    & - &(k.p_W)M^4 \Big ]
\end{eqnarray*}

\begin{eqnarray*}
\chi_2  = & - &
  20(p.p_b)(p.k)(k.q)  + 12(p.p_b)(p.k)^2
+  8(p.p_b)(k.q)^2 - (p.p_b)M^2 q^2
\\
+\mfrac{1}{(p.k)} &\Big [&  4(p.q)(p_b.k) - 4(p.q)M^2 - 2(p_b.k) q^2
  - 6(p.k)(p_b.k) m_t^2 + 4(p_b.q)(k.q)
\\
& - &  3M^2 m_t^2 - 4(p.k)(p_b.q) \Big]
\\
& +& 2(p.q)(p_b.q)M^2 + 4(p.q)(k.q)M^2
+ 8(p_b.k)(k.q) m_t^2  +  6(p_b.q) M^2 m_t^2
+ 7(k.q) M^2 m_t^2  
\\
+\mfrac{1}{M_W^2} 
&\Big [ &24(p.p_b)(p.k)(p.p_W)(k.p_W)
- 24(p.p_b)(p.k)(k.p_W)(q.p_W) + 16(p.p_b)(p.q)(k.p_W)^2
\\
&- & 16(p.p_b)(p.p_W)(k.q)(k.p_W)
+ 4(p.p_b)(p.p_W)(q.p_W)M^2 + 16(p.p_b)(k.q)(k.p_W)(q.p_W)
\\
&-& 8(p.k)(p.p_W)(p_b.q)(k.p_W) + 2(p.k)(p.p_W)(q.p_W)  M^2
- 8(p.p_b)(k.p_W)^2 q^2
\\
&-& 4(p.p_b)(q.p_W)^2  M^2 + 8(p.k)(p.q)(p_b.p_W)(k.p_W)
 + 8(p.k)(p_b.q)(k.p_W)(q.p_W)
\\
& -& 4(p.k)(p_b.p_W)(k.p_W) q^2  - 2(p.k)(q.p_W)^2  M^2
+ 4(p.q)(p.p_W)(p_b.p_W)M^2
\\
&+& 2(p.q)(p.p_W)(k.p_W)M^2  +   6(p.q)(k.p_W)(q.p_W) M^2
 + 4(p.p_W)(p_b.q)(q.p_W)M^2
\\
&-& 2(p.p_W)(p_b.p_W) M^2q^2 + 2(p.p_W)(k.q)(q.p_W)  M^2
- 4(p.p_W)(k.p_W) M^2 q^2
\\
& - & 4(p.p_W)^2(p_b.q) M^2 -  2(p.p_W)^2 (k.q)  M^2 \Big]
  \\
+ \mfrac{2m_t^2}{M_W^2} &\Big[ & - 6(p.p_W)(p_b.k)(k.p_W) 
- 3(p.p_W)(k.p_W) M^2 
+ 6(p_b.k)(k.p_W)(q.p_W)  
\\
&&- 2(p_b.q)(k.p_W)^2  
+ 2(p_b.p_W)(k.q)(k.p_W)   - 2(p_b.p_W)(q.p_W)  M^2 
+ 2(k.p_W)(q.p_W)  M^2  \Big]
\end{eqnarray*}
\begin{eqnarray*}
 \chi_{int}   = & - &         
 8(p.p_b)(p.k)(p_b.k) + 4(p.p_b)(p.k)M^2
 - 4(p.p_b)(p_b.k)(k.q) - 8(p.p_b)(p_b.q) M^2
\\
&  + & 4(p.q)(p_b.k)^2
 +  8(p.p_b)^2 M^2 - 4(p.k)^2 M^2  + 6(p.q)(p_b.k) M^2
 \\
 & +&  12(p.k)(p_b.k)(p_b.q)
  +   6(p.k)(p_b.q) M^2 + 4(p.k)(k.q)M^2
  + 2(p_b.k) M^2 m_t^2
\\
&  +  & 4(p_b.k)^2 m_t^2 - (p.q) M^4
 -  10(p.p_b)(k.q)M^2
 \\
  + \mfrac{4(p.p_b)}{M_W^2} & \Big[&
 - 2 (p.p_W)(p_b.k)(k.p_W)
 + 4(p.p_W)(p_b.p_W)M^2
 + 4(p.p_W)(k.p_W)  M^2
\\
&& + 2 (p_b.k)(k.p_W)(q.p_W) 
- 2 (p_b.p_W)(q.p_W)M^2
-  (k.p_W)(q.p_W)  M^2 \Big]
 \\
 +  \mfrac{4}{M_W^2}  &\Big[&   (p.k)(k.p_W)(q.p_W)  M^2 
-  2(p.q)(p_b.k)(p_b.p_W)(k.p_W)
 + (p.q)(p_b.p_W)(k.p_W)  M^2
\\
&& +  2(p.q)(p_b.p_W)^2  M^2 - (p.q)(k.p_W)^2  M^2 \Big]
\\
 + \mfrac{4(p.p_W)}{M_W^2} & \Big[& 
- 2(p.k)(p_b.k)(p_b.p_W)  - 2(p.k)(p_b.p_W)  M^2
- 2(p.k)(k.p_W)  M^2 
+ 2(p_b.k)(p_b.p_W)(k.q)
\\
&& -  (p_b.k)(q.p_W)  M^2 - 2(p_b.k)^2(q.p_W) 
- 2(p_b.q)(p_b.p_W)  M^2
- 3(p_b.q)(k.p_W)  M^2
\\
&& +  2(p_b.p_W)(k.q)  M^2
+ (k.q)(k.p_W)  M^2
+ (p.p_W) (p_b.k)  M^2 + 2(p.p_W)(p_b.k)^2  \Big]
\end{eqnarray*}